\documentclass[preprint,showpacs,amsmath,amssymb,prd]{revtex4-1}
\usepackage{epsf}
\usepackage{amsmath,amsthm,amssymb}
\usepackage{color}
\usepackage{dcolumn}
\usepackage{bm}
\usepackage{graphicx,epsfig}
\usepackage[dvipsnames]{xcolor}

\graphicspath{{fig/}}
\everymath{\displaystyle}

\begin{document} 

\title{Paraxial wave function and Gouy phase for a relativistic 
electron in a uniform magnetic field}

\author{Liping Zou$^{1}$}
\email{zoulp@impcas.ac.cn}
\author{Pengming Zhang$^{2}$}
\email{zhangpm5@mail.sysu.edu.cn}
\author{Alexander J. Silenko$^{1,3,4}$}
\email{alsilenko@mail.ru}

\affiliation{$^1$Institute of Modern Physics, Chinese Academy of
Sciences, Lanzhou 730000, China}
\affiliation{$^2$School of Physics and Astronomy, Sun Yat-sen University, Zhuhai 519082, China}
\affiliation{$^3$Bogoliubov Laboratory of Theoretical Physics, Joint
Institute for Nuclear Research, Dubna 141980, Russia}
\affiliation{$^4$Research Institute for Nuclear Problems, Belarusian
State University, Minsk 220030, Belarus}

\date{\today}

\begin{abstract}
A connection between relativistic quantum mechanics in the Foldy-Wouthuysen 
representation and the paraxial equations is established for a Dirac particle 
in external fields.
The paraxial form of the Landau eigenfunction for a relativistic electron 
in a uniform magnetic field is determined. The obtained wave function contains 
the Gouy phase and significantly approaches to the paraxial wave function for a free electron.
\end{abstract}

\maketitle

The prediction \cite{Bliokh2007} and discovery \cite{UTV} of twisted (vortex) 
electrons in a free space conditions an importance of a detailed quantum-mechanical 
description of such particles. For this purpose, the paraxial equation is mostly 
applied. The approach based on the paraxial equation is widely used in optics 
for studying twisted and untwisted structured light beams 
\cite{Siegman,Allen,TwPhotons2,TwPhotRev3,BBP}. The connection of this approach 
with traditional approaches of relativistic quantum mechanics (QM) is considered, 
e.g., in Refs. \cite{Barnett-FWQM,BliokhSOI,LightArXiv}.

In contemporary studies of twisted electrons, an important place is occupied by 
their interactions with a magnetic field \cite{Bliokhmagnetic,magnetic,Kruining,Rajabi,Greenshields,classicalmagnetic,
experimentmagnetic,Rusz,Edstrom,Observation,Manipulating,ResonanceTwistedElectrons,
PhysRevLettEQM2019,snakelike}.
A similarity between the wave function for a free twisted electron and the 
Landau wave function for an electron in a uniform magnetic field is evident and 
was much discussed (see Refs. \cite{BliokhSOI,Bliokhmagnetic,Lloyd,Larocque}). 
However, approaches used in the two cases substantially differ. The Landau 
solution \cite{Fock,Landau,LL3} has been obtained in the framework of nonrelativistic 
Schr\"{o}dinger-Pauli QM. The free twisted electron is described by the paraxial 
equation.  The connection between the relativistic QM and the paraxial equation 
has been analyzed in Refs. \cite{Barnett-FWQM,photonPRA}. To establish this 
connection, it is convenient to use the Foldy-Wouthuysen (FW) representation 
\cite{FW}. In this representation, relativistic QM takes the form equivalent 
to nonrelativistic Schr\"{o}dinger QM. The Hamiltonian
and all operators in the FW representation are even, i.e., block diagonal 
(diagonal in two spinors). Relations between the operators in this representation 
are similar to those between the respective
classical quantities. The form of quantum-mechanical operators
for relativistic particles in external fields is the same
as in the nonrelativistic quantum theory. In particular, the operators of the 
position and momentum
are equal to $\bm r$ and $\bm p = -i\hbar\nabla$, respectively \cite{FW,Costella,dVFor,JMP,relativisticFW,Reply2019,SpinDiracFWR}.

The exact relativistic Hamiltonian in the FW representation describing the Dirac 
electron in the uniform magnetic field $\bm B=B\bm e_z$ is defined by 
\cite{Case,Energy1,JMP,Energy3}
\begin{equation}
\begin{array}{c}
i\frac{\partial\Psi_{FW}}{\partial t}={\cal H}_{FW}\Psi_{FW},\qquad {\cal H}_{FW}=\beta\sqrt{m^2+\bm{\pi}^2-e\bm\Sigma\cdot\bm B},
\end{array}
\label{eq33new}
\end{equation}
where $\bm{\pi}=\bm{p}-e\bm A$ is the kinetic momentum and $\beta$ and $\bm\Sigma$ are 
the Dirac matrices.
This Hamiltonian acts on the bispinor $\Psi_{FW}=
\left(\begin{array}{c} \Phi_{FW} \\ 0 \end{array}\right)$. In the present study, we use the 
system of units $\hbar=1,~c=1$. We include $\hbar$ and $c$ explicitly when this inclusion
clarifies the problem.

Since eigenfunctions of the FW Hamiltonian (\ref{eq33new}) are also eigenfunctions 
of the operator $\bm{\pi}^2$, they are defined by the nonrelativistic Landau solution \cite{Case,Energy1,Energy3}. The eigenfunctions are more complicated when the Dirac 
representation is used \cite{Kruining,Rajabi,Energy2a,Energy2,OConnell,CanutoChiu}. 
Certainly, the energy eigenvalues
do not depend on a representation and are given by \cite{Kruining,Rajabi,Case,Energy1,JMP,Energy3,Energy2a,Energy2,OConnell,CanutoChiu}
\begin{equation}
\begin{array}{c}
E=\sqrt{m^2+p_z^2+(2n+1+|\ell|+\ell+2s_z)|e|B},
\end{array}
\label{eqOAM}
\end{equation} where $n=0,1,2,\dots$ is the radial quantum number and $\ell$ 
is an eigenvalue 
of the orbital angular momentum (OAM) operator projected on the $z$ axis, 
$\ell=l_z=(\bm r\times\bm p)_z$. In the considered case, $A_\phi=Br/2,\,A_r=A_z=0,\,e=-|e|$.
The relativistic approach (unlike the nonrelativistic one) demonstrates that the 
Landau levels are not
equidistant for any field strength \cite{ResonanceTwistedElectrons}. Amazingly, 
eigenfunctions (more precisely, upper spinors) of the \emph{relativistic} 
FW Hamiltonian are defined by the \emph{nonrelativistic} Landau solution (see Refs. \cite{Case,Energy1,Energy3}). In the cylindrical coordinates, these eigenfunctions 
are the Laguerre-Gauss beams:
\begin{equation}
\begin{array}{c}
\Phi_{FW}={\cal A}\exp{(i\ell\phi)}\exp{(ip_zz)},\qquad \int{\Phi_{FW}^\dag\Phi_{FW} rdrd\phi}=1,\\
{\cal A}=\frac{C_{n\ell}}{w_m}\left(\frac{\sqrt2r}{w_m}\right)^{|\ell|}
L_n^{|\ell|}\left(\frac{2r^2}{w^2_m}\right)\exp{\left(-\frac{r^2}{w^2_m}\right)}\eta,\\
C_{n\ell}=\sqrt{\frac{2n!}{\pi(n+|\ell|)!}},\qquad
w_m=\frac{2}{\sqrt{|e|B}},
\end{array}
\label{Lenergy}
\end{equation} where the real function ${\cal A}$ defines the amplitude of the beam, 
and $L_n^{|\ell|}$ is the generalized Laguerre polynomial. Since the spin operator in 
the FW representation, $\bm s=\hbar\bm\Sigma/2$, commutes with the Hamiltonian (\ref{eq33new}) 
and the zero lower spinor of the bispinor $\Psi_{FW}$ is disregarded, the spin 
function $\eta$ is an eigenfunction of the Pauli operator $\sigma_z$ (cf. Ref. \cite{Energy3}):
$$\sigma_z\eta^+=\eta^+,\quad \sigma_z\eta^-=-\eta^-,\quad
\eta^+=\left(\begin{array}{c} 1 \\ 0 \end{array}\right),\quad \eta^-=\left(\begin{array}{c} 0 \\ 1 \end{array}\right).$$

It is important that the solution (\ref{Lenergy}) can be obtained from the exact relativistic
wave function in Dirac representation \cite{Rajabi} by the use of the connection
between the Dirac and FW wave functions found in Ref. \cite{Energy3}.

The electron possesses a small \emph{anomalous} magnetic moment which is not 
taken into account in Eqs. (\ref{eqOAM}), (\ref{Lenergy}). Due to its existence, 
a consideration of the spin does not lead to an additional degeneracy of the 
energy levels. The solution (\ref{Lenergy}) does not contain the Gouy phase.

The use of the FW representation is necessary to connect 
relativistic quantum-mechanical equations and paraxial ones. The latter 
equations can be introduced when the paraxial approximation $|\bm p_\bot|\ll p$ 
is satisfied. Operators entering these equations are equivalent to the 
corresponding operators of Schr\"{o}dinger QM. Since the FW representation 
restores the Schr\"{o}dinger picture of relativistic QM \cite{FW,Costella,dVFor,relativisticFW,Reply2019,SpinDiracFWR,JINRLett12}, one 
needs to use FW Hamiltonians. For the considered problem, the Hamiltonian (\ref{eq33new}) 
is exact. In other cases, approximate relativistic FW Hamiltonians can be derived 
by various methods \cite{JMP,relativisticFW,PRA2008,TMPFW,PRA2015,ChiouChen,PRAExpO}.

Similarly to Refs. \cite{Barnett-FWQM,LightArXiv}, we can determine a connection 
between the relativistic quantum-mechanical equations and paraxial ones for a 
particle \emph{in external fields}. In Refs. \cite{Barnett-FWQM,LightArXiv}, 
particles in a free space have been considered.
For stationary states, ${\cal H}_{FW}\Psi_{FW}=E\Psi_{FW}$. Let us denote $P=\sqrt{E^2-m^2}=\hbar k$.
Squaring
Eq. (\ref{eq33new}) for the upper spinor and applying the paraxial approximation for
$p_z>0$ results in (cf. Refs. \cite{Barnett-FWQM,LightArXiv})
\begin{equation}\begin{array}{c}
P^2=\bm\pi^2-e\bm\Sigma\cdot\bm B=\bm\pi_\bot^2+p_z^2-e\bm\Sigma\cdot\bm B,\\
p_z=\sqrt{P^2-\bm\pi_\bot^2+e\bm\Sigma\cdot\bm B}\approx P-\frac{\bm\pi_\bot^2-e\bm\Sigma\cdot\bm B}{2P}.
\end{array}
\label{eqnnumb}
\end{equation}
This transformation leads to the following equation:
\begin{equation}
\begin{array}{c}
\left(\bm\pi_\bot^2-e\bm\Sigma\cdot\bm B+2Pp_z\right)\Phi_{FW}=2P^2\Phi_{FW}.
\end{array}
\label{eqpqm}
\end{equation}
An equivalent form of this equation reads
\begin{equation}
\begin{array}{c}
\left(\bm\pi_\bot^2-e\bm\Sigma\cdot\bm B-2ik\frac{\partial}{\partial
z}\right)\Phi_{FW}=2k^2\Phi_{FW},\qquad
\bm\pi_\bot^2=-\nabla^2_\bot+ieB\frac{\partial}{\partial\phi}+\frac{e^2B^2r^2}{4},
\\ \nabla^2_\bot=
\frac{\partial^2}{\partial r^2}+\frac1r\frac{\partial}{\partial
r}+\frac{1}{r^2}\frac{\partial^2}{\partial\phi^2}.
\end{array}
\label{eqpqmef}
\end{equation}
Equation (\ref{eqpqmef}) is an approximate form of the general equation (\ref{eq33new}) 
when the paraxial approximation is satisfied.
The substitution $\Phi_{FW}=\exp{(ikz)}\Psi$ brings the corresponding paraxial 
equation
\begin{equation}
\begin{array}{c}
\left(\nabla^2_\bot-ieB\frac{\partial}{\partial\phi}
-\frac{e^2B^2r^2}{4}+2es_zB+2ik\frac{\partial}{\partial
z}\right)\Psi=0.
\end{array}
\label{eqpar}
\end{equation} This substitution shifts the squared particle momentum and is 
equivalent to a shift of the zero energy level in Schr\"{o}dinger QM.
Within the paraxial approximation, Eq. (\ref{eqpar}) properly describes electrons of
arbitrary energies in a uniform magnetic field. Therefore, the use of the FW 
transformation establishes a
connection between relativistic QM and the respective relativistic paraxial 
equation. We
underline the difference between $\Phi_{FW}$ and $\Psi$.

The Landau solution defines the eigenvalues of the operator describing the 
transversal motion:
\begin{equation}
\begin{array}{c}
\left(\nabla^2_\bot-ieB\frac{\partial}{\partial\phi}-\frac{e^2B^2r^2}{4}+2es_zB\right)\Phi_{FW}
=-\left(2n+1+|\ell|+\ell+2s_z\right)|e|B\Phi_{FW}.
\end{array}
\label{botmt}
\end{equation}  The same equation can be written for the paraxial wave 
function $\Psi$. This equation allows us to determine the Gouy phase.

The Landau wave function contains the exponential factor $\exp{[i(p_z/\hbar)z]}$.
Taking into account the connection between $\Phi_{FW}$ and $\Psi$, we obtain 
that the latter wave function is proportional to the exponential factor
\[
\exp{\left(i\frac{p_z}{\hbar}z\right)}\exp{(-ikz)}
=\exp{\left(-i\frac{P-p_z}{\hbar}z\right)}.
\]
Equations (\ref{Lenergy}), (\ref{eqnnumb}), and (\ref{botmt}) result in the 
following form of the paraxial wave function:
\begin{equation}
\begin{array}{c}
\Psi={\cal A}\exp{(i\ell\phi)}\exp{[-i\zeta(z)]},
\qquad \int{\Psi^\dag\Psi rdrd\phi}=1,\\
\zeta=\left(2n+1+|\ell|+\ell+2s_z\right)\frac{|e|B}{2k}z
=\left(2n+1+|\ell|+\ell+2s_z\right)\frac{2z}{kw_m^2},
\end{array}
\label{Lparaxl}
\end{equation} where $\zeta$ is the Gouy phase.
Evidently, this wave function satisfies the paraxial equation (\ref{eqpar}).

Equation (\ref{Lparaxl}) shows that the wave eigenfunction of the relativistic 
electron in the uniform magnetic field acquires the Gouy phase $\zeta$ after the 
transition to the paraxial equation. This property increases the similarity 
between the wave eigenfunctions of the relativistic Landau electron in the 
uniform magnetic field and the twisted electron in a free space. In the latter 
case, the eigenfunction reads \cite{BliokhSOI,LightArXiv,Bliokhmagnetic}
\begin{equation}
\begin{array}{c}
\Psi=\mathbb{A}\exp{(i\ell\phi)}\exp{\left[i\frac{kr^2}{2R(z)}\right]}\exp{[-i\zeta(z)]},\\
\mathbb{A}=\frac{C_{n\ell}}{w(z)}\left(\frac{\sqrt2r}{w(z)}\right)^{|\ell|}
L_n^{|\ell|}\left(\frac{2r^2}{w^2(z)}\right)\exp{\left(-\frac{r^2}{w^2(z)}\right)}\eta,\\
C_{n\ell}=\sqrt{\frac{2n!}{\pi(n+|\ell|)!}},\quad w(z)=w_0\sqrt{1+\frac{z^2}{z_R^2}},\quad
R(z)=z+\frac{z_R^2}{z},\\
\zeta(z)=(2n+|\ell|+1)\arctan{\left(\frac{z}{z_R}\right)},\quad z_R=\frac{kw_0^2}{2},
\end{array}
\label{eqnew}
\end{equation}
where $w_0$ is the minimum beam width, $R(z)$ is the radius of curvature of the 
wavefront, and $z_R$ is the Rayleigh diffraction length.

An analysis of Eqs. (\ref{Lenergy}), (\ref{Lparaxl}), and (\ref{eqnew}) shows 
the substantial similarity between the paraxial wave functions of relativistic 
Dirac electrons in a uniform magnetic field and a free space. The former case 
(Landau solution) characterizes the wave with the infinite radius of curvature 
of the wavefront, $R(z)\rightarrow\infty$. In this case, 
$z_R\rightarrow\infty, ~\arctan{(z/z_R)}\approx z/z_R$, and the paraxial wave 
functions becomes equivalent provided that $w_0=w_m$.

It can be similarly obtained that the paraxial wave function of relativistic 
positrons ($e=|e|$) in a uniform magnetic field has the form (\ref{Lparaxl}) 
where the Gouy phase differs by signs:
\[
\zeta=\left(2n+1+|\ell|-\ell-2s_z\right)\frac{|e|B}{2k}z
=\left(2n+1+|\ell|-\ell-2s_z\right)\frac{2z}{kw_m^2}.
\]
The Landau solution describes a motion of a charged particle in a uniform 
magnetic field. This motion is governed by the Lorentz force and a direction 
of a particle rotation is definite. A simple analysis shows that 
\emph{physically correct} solutions of the related quantum-mechanical 
equations correspond to $\ell\ge0$ for electrons and $\ell\le0$ for positrons.

The obtained connection between relativistic QM in the FW representation and 
the paraxial equations can be applied to a wide class of problems connected 
with relativistic electron (positron, muon) beams in different external fields \cite{LL4,Glushkov,Ihra,Rao,Gottwald,AVGlushkov,ChavezLBM} (e.g., crossed 
magnetic and electric fields).

In summary, we have presented the first 
attempt to establish a connection between relativistic QM in the FW representation 
and the paraxial equations for a
Dirac particle in external fields. For a relativistic electron in a uniform 
magnetic field, the paraxial form of the Landau eigenfunction contains the 
Gouy phase and amazingly approaches to the paraxial wave function for a free 
electron. The Gouy phase does not enter the standard quantum-mechanical solutions 
in the Dirac and FW representations. We have demonstrated for the first time that
it appears as a result of a transition from QM in the FW representation 
(or Schr\"{o}dinger-Pauli QM) to the paraxial quantum-mechanical equation.

\begin{acknowledgments}
This work was supported by the Belarusian Republican Foundation
for Fundamental Research
(Grant No. $\Phi$18D-002), by the National Natural Science
Foundation of China (Grants No. 11575254 and No. 11805242), and
by the National Key Research and Development Program of China
(No. 2016YFE0130800).
A. J. S. also acknowledges hospitality and support by the
Institute of Modern
Physics of the Chinese Academy of Sciences. The authors are
grateful to
I. P. Ivanov for helpful exchanges.
\end{acknowledgments}

\end{document}